\documentclass[12pt]{article}
\usepackage{amssymb}
\usepackage{amsmath}
\usepackage{apacite}
\usepackage{algorithm}
\usepackage{algpseudocode}
\usepackage{latexsym}
\usepackage{epsfig}
\usepackage{graphicx}
\usepackage{subfigure}
\usepackage{wasysym}
\usepackage{threeparttable}
\usepackage{natbib}
\usepackage{color}
\usepackage{epstopdf}
\usepackage{bm}
\usepackage{float}
\usepackage{todonotes}
\usepackage{verbatim}
\usepackage{svg}
\usepackage{amsmath}
\usepackage[utf8]{inputenc}

\usepackage{hyperref}

\def\log{\hbox{log}}

\def\boxit#1{\vbox{\hrule\hbox{\vrule\kern6pt
          \vbox{\kern6pt#1\kern6pt}\kern6pt\vrule}\hrule}}

\def\bse{\begin{eqnarray*}}
\def\ese{\end{eqnarray*}}
\def\be{\begin{eqnarray}}
\def\ee{\end{eqnarray}}
\def\bq{\begin{equation}}
\def\eq{\end{equation}}
\def\bse{\begin{eqnarray*}}
\def\ese{\end{eqnarray*}}

\begin{document}

\thispagestyle{empty} 
\baselineskip=28pt

\begin{center}
{\LARGE{\bf Scalable Joint Modeling of Dependent Multi-Type Survey Data for Small Area Estimation}}
\end{center}

\baselineskip=12pt

\vskip 2mm
\begin{center}
Zewei Kong\footnote{(\baselineskip=10pt to whom correspondence should be addressed) Department of Statistics and Data Science, University of Missouri,
146 Middlebush Hall, Columbia, MO 65211-6100, zk3bx@mail.missouri.edu},
Paul A. Parker\footnote{\baselineskip=10pt Department of Statistics, University of California Santa Cruz, 1156 High Street, Santa Cruz, CA 95064,
paulparker@ucsc.edu},\,
    and  Scott H. Holan\footnote{\baselineskip=10pt Department of Statistics and Data Science, University of Missouri
    146 Middlebush Hall, Columbia, MO 65211-6100, holans@missouri.edu}
\\
\end{center}
%
%
%
%
\vskip 4mm
\baselineskip=12pt 
\begin{center}
{\bf Abstract}
\end{center}
We develop a Bayesian area-level small area estimation framework that jointly models binomial and Gaussian survey responses through shared spatial random effects. This work is motivated by the American Community Survey (ACS), which provides useful information that contributes to federal funding and policy making decisions, and often yields direct estimates with large standard errors in small domains. The proposed Multi-type model borrows strength across outcomes and spatial neighbors to improve the precision of the associated estimates. For the binomial component, P\'olya--Gamma data augmentation yields a conditionally Gaussian representation, while spatial basis functions provide dimension reduction for high-dimensional spatial data. Together, these features lead to closed-form conditional posteriors and, thus, an efficient Gibbs sampler. Through empirical simulations, we show that the proposed joint model improves estimation precision relative to independent Univariate models. Applying the method to ACS median income and poverty rate data, we find that the proposed Multi-type model yields similar point estimates but smaller posterior variances than the corresponding Univariate models.

\baselineskip=12pt
\par\vfill\noindent
\par\medskip\noindent
\clearpage\pagebreak\newpage \pagenumbering{arabic}
\baselineskip=24pt

\section{Introduction}\label{sec: intro}

Small area estimation (SAE) is  important in the context of official statistics for producing estimates in domains with limited sample sizes. In these sparse domains, direct survey estimates typically have large design-based variances. Model-based SAE improves estimation precision by leveraging dependence and combining direct estimates with auxiliary information and cross-domain structure. This reduces the variability of the domain-level estimates and provides associated measures of uncertainty \citep{rao2015small}.

For area-level estimation, the Fay--Herriot (FH) model \citep{fay1979estimates} serves as the foundational framework. By treating the sampling variances of the direct estimates as known and incorporating them into the model, the FH framework allows the resulting estimates to reflect the precision of the direct estimates. Since then, the area-level SAE framework has been extended in several directions. {{These including spatial FH models \citep{porter2014spatial}}}, multivariate FH models \citep{porter2015small,benavent2016multivariate}, spatio-temporal FH models \citep{marhuenda2013small}, and Bayesian modeling under more complex heteroskedastic structures \citep{parker2024conjugate}.

However, these developments have mostly focused on settings where the responses are of the same type, most often Gaussian. Within official statistics, in practice, non-Gaussian responses or multiple types of variables are often of interest. For example, the Small Area Income and Poverty Estimates (SAIPE) program requires precise estimates for both median household income, which is continuous, and {{poverty rates, which are aggregates of the underlying poverty status indicator.}} Historically, these outcomes have often been modeled separately \citep{franco2013applying}. Recent work has considered joint modeling for mixed-type outcomes \citep{ekvall2022mixed,sun2024multivariate}. {{For example, \citet{ekvall2022mixed} linked responses of different types through a latent Gaussian framework. We develop a joint Bayesian hierarchical model that directly models Gaussian and binomial responses through their respective likelihoods and links them through shared spatial random effects.}}

At the same time, building such a joint Bayesian model is often computationally demanding. Combining non-Gaussian data with spatial random effects typically breaks conditional conjugacy. As a result, posterior inference often relies on non-conjugate Markov chain Monte Carlo (MCMC) steps such as Metropolis--Hastings, which can become a computational bottleneck when the number of areas is large. Recent work in SAE has addressed this issue by developing scalable Bayesian methods with conjugate structure. For example, \citet{bradley2020bayesian} developed Bayesian hierarchical models with conjugate full conditional distributions for dependent data from the natural exponential family, while \citet{parker2021general} proposed a fully conjugate Bayesian model for heteroskedastic survey data. 

Motivated by these advances, we propose a scalable Bayesian area-level joint framework that links Gaussian and binomial outcomes through shared spatial random effects. The framework combines three components to address the challenges of non-conjugacy, high-dimensional spatial effects, and design-based variance inputs. First, we apply P\'olya--Gamma data augmentation \citep{polson2013bayesian} to transform the logistic likelihood into a conditionally Gaussian form. This yields closed-form updates and enables efficient Gibbs sampling. {{Second, we use spatial basis functions \citep{hughes2013dimension,bradley2015multivariate} to model the spatial random effects in a lower-dimensional form while preserving the main spatial dependence structure.}} Third, we use an effective sample size (ESS) approximation \citep{chen2014use} to the variance. This allows the binomial model variance structure to reflect the design-based variances derived from ACS margins of error.

In summary, this paper makes three main contributions. First, we develop a scalable area-level joint spatial framework for Gaussian and binomial survey data. Second, by combining P\'olya--Gamma augmentation with a low-rank spatial structure, we obtain a computationally efficient Gibbs sampler with closed-form conditional updates. Third, through simulation studies and an ACS application, we show that the joint model yields smaller posterior variances than separate Univariate models while preserving similar posterior mean surfaces, leading to improved uncertainty quantification. The remainder of this paper is organized as follows. Section~\ref{sec:methods} introduces the proposed Bayesian framework. Section~\ref{sec:simulation} evaluates the proposed methodology through simulation studies. {{Section~\ref{sec:data} applies our model to 2021 ACS tract-level 5-year period estimates to jointly estimate area-level median household income and poverty rates.}} Finally, Section~\ref{sec:conclusion} provides concluding discussion.

\section{Multi-type Model}\label{sec:methods}
We jointly model area-level Gaussian and binomial survey outcomes in a Bayesian hierarchical framework, linking the two responses through a shared low-rank spatial random effect. Conditional on the latent effects, the Gaussian and binomial components are independent. P\'olya-Gamma data augmentation transforms the binomial likelihood into a conditionally Gaussian form, which leads to straightforward Gibbs updates.

\subsection{Univariate spatial models}

Let $N$ denote the total number of geographic areas. As baselines, we fit separate spatial models to Gaussian and binomial responses. The Gaussian and binomial outcomes are modeled independently using distinct spatial coefficient vectors, denoted $u_1$ and $u_2$. Dimension reduction for the spatial random effects is achieved using spatial basis functions. Let $A$ denote the $N \times N$ binary adjacency matrix of the areal graph, and let
\[
A = U \Lambda U^\top
\]
be its eigendecomposition, where $U$ contains the orthonormal eigenvectors and $\Lambda$ is the diagonal matrix of eigenvalues. We retain the eigenvectors associated with the positive eigenvalues and use them as spatial basis functions; see \cite{bradley2016bayesian} for a related reduced-rank construction. Related adjacency-eigenvector basis representations are also used by \cite{parker2022computationally}. {{Let $J$ denote the retained index set, with cardinality $r = |J|$. Let $\Phi \in \mathbb{R}^{N \times r}$ denote the matrix formed by the retained eigenvectors, where $\phi_i^\top$ denotes the $i$-th row of $\Phi$. The vector $\phi_i \in \mathbb{R}^r$ is the spatial basis vector for area $i$.}}

For the Gaussian response, let $Y_{1i}$ denote the direct estimate on the original scale for area $i$. We apply a log transformation and then standardize the transformed values so that the Gaussian response is on a scale comparable to the logit-scale binomial component. Let $Z_{1i}$ denote the resulting standardized value, and let $V_{1i}$ denote the corresponding estimated design-based variance on the standardized log scale, treated as known in the model.

The area-level Gaussian model is specified as
\begin{equation*}
Z_{1i} \mid \mu_{1i} \sim N(\mu_{1i}, V_{1i}), \quad
\mu_{1i} = x_{1i}^\top \beta_1 + \phi_i^\top u_1 + \zeta_{1i},
\end{equation*}
for $i=1,\dots,N$. Here, $x_{1i} \in \mathbb{R}^{p_1}$ is the covariate vector, $u_1 \in \mathbb{R}^r$ is the Gaussian-specific spatial coefficient vector, and $\zeta_{1i}$ is an unstructured area-specific random effect. {{We assign independent priors $\beta_1 \sim N(0, \sigma_\beta^{2} I_{p_1})$ and $u_1 \mid \sigma_{u_1}^2 \sim N(0, \sigma_{u_1}^2 I_r)$. We set $\sigma_\beta^{2} = 1000$ throughout this paper, corresponding to a weakly informative Gaussian prior. }} We also assume $\zeta_1 \mid \sigma_{\zeta_1}^2 \sim N(0, \sigma_{\zeta_1}^2 I_N)$, where $\zeta_1 = (\zeta_{11}, \dots, \zeta_{1N})^\top$. Variance components are assigned priors $\sigma_{u_1}^2 \sim \text{InvGamma}(a_u, b_u)$ and $\sigma_{\zeta_1}^2 \sim \text{InvGamma}(a_{\zeta_1}, b_{\zeta_1})$.

For the binomial response, let $\hat p_i$ denote the direct survey estimate of the rate and let $\widehat{\mathrm{Var}}(\hat p_i)$ denote its design-based variance. Following \cite{chen2014use}, we define the effective sample size
\[
m_i^E = \frac{\hat p_i(1-\hat p_i)}{\widehat{\mathrm{Var}}(\hat p_i)},
\]
and the corresponding effective count
\[
Z_{2i}^E = m_i^E \hat p_i.
\]
{{We use a binomial likelihood based on the effective sample size construction of \citet{chen2014use}. Here, $\pi_i$ denotes the area-level mean (i.e., the true population proportion), with $\operatorname{logit}(\pi_i)$ given by the linear predictor. }}
Specifically,

\[
L_i(\pi_i \mid Z_{2i}^E,m_i^E)\propto
\pi_i^{Z_{2i}^E}(1-\pi_i)^{m_i^E-Z_{2i}^E},
\]
with
\[
\mathrm{logit}(\pi_i)=x_{2i}^\top\beta_2+\phi_i^\top u_2+\zeta_{2i}.
\]

The vector $u_2 \in \mathbb{R}^r$ is the binomial-specific spatial coefficient vector and $\zeta_{2i}$ is the unstructured random effect. We use analogous priors for the binomial block: $\beta_2 \sim N(0, \sigma_\beta^{2} I_{p_2})$, $u_2 \mid \sigma_{u_2}^2 \sim N(0, \sigma_{u_2}^2 I_r)$, and $\zeta_2 \mid \sigma_{\zeta_2}^2 \sim N(0, \sigma_{\zeta_2}^2 I_N)$. The variance components are assumed to have independent $\text{InvGamma}(a_u, b_u)$ and $\text{InvGamma}(a_{\zeta_2}, b_{\zeta_2})$ priors.

\subsection{Multi-type Response Model}

The separate Univariate models assume independence and therefore fail to borrow strength across the correlated survey responses. To capture cross-response dependence, we introduce a joint Multi-type model with a shared spatial random effect.

The joint model incorporates the spatial basis matrix $\Phi$ into a shared hierarchical structure
\begin{align*}
Z_{1i} \mid \mu_{1i} &\sim N(\mu_{1i}, V_{1i}), \quad \mu_{1i} = x_{1i}^\top \beta_1 + \tau_1 \phi_i^\top \eta + \zeta_{1i}, \\
L_i(\pi_i \mid Z_{2i}^E,m_i^E) &\propto
\pi_i^{Z_{2i}^E}(1-\pi_i)^{m_i^E-Z_{2i}^E},
\,\,\,\,
\text{logit}(\pi_i)=x_{2i}^\top\beta_2+\tau_2\phi_i^\top\eta+\tau_3\phi_i^\top\kappa+\zeta_{2i}.
\end{align*}
The vector $\eta \in \mathbb{R}^r$ denotes the shared spatial effect that induces dependence between the two outcomes, while $\kappa \in \mathbb{R}^r$ captures spatial variation specific to the binomial outcome. The coefficients $\tau_1$ and $\tau_2$ scale the shared spatial effect in the Gaussian and binomial components, respectively, and $\tau_3$ scales the binomial-specific spatial effect. The unstructured area effects $\zeta_{1i}$ and $\zeta_{2i}$ capture non-spatial residual heterogeneity.

We assume Gaussian priors for the fixed effects $\beta_l \sim N(0, \sigma_\beta^{2} I_{p_l})$ for $l=1,2$. To avoid confounding between the scale of the latent spatial vectors and the coefficients $\tau_j$, we assign standard normal priors $\eta \sim N(0, I_r)$ and $\kappa \sim N(0, I_r)$. The coefficients $\tau_j$ are assigned $\tau_j \sim N(0, \sigma_\tau^{2})$ for $j=1,2,3$. {{We set $\sigma_\tau^{2} = 100$, corresponding to a vague prior for $\tau_j \sim N(0,100)$ for $j=1,2,3$. }} We further assume $\zeta_l \mid \sigma_{\zeta_l}^2 \sim N(0, \sigma_{\zeta_l}^2 I_N)$ and $\sigma_{\zeta_l}^2 \sim \text{InvGamma}(a_l, b_l)$ for $l=1,2$.

By representing the spatial random effects in the reduced basis space defined by $\Phi$, the dominant matrix inversions in the MCMC updates are reduced from dimension $N$ to dimension $r$, so the main computational cost scales with $O(r^3)$ rather than $O(N^3)$.
Table~\ref{tab:symbols} summarizes the primary notation used in the joint model framework.

\begin{table}[ht]
\centering
\caption{Symbols in the joint model and their interpretation.}
\label{tab:symbols}
\begin{tabular}{ll}
\hline
Symbol & Meaning \\
\hline
$Z_{1i}$, $Z_{2i}^E$ & Standardized log-transformed direct estimate and effective count for area $i$\\
$m_i^E$ & Effective sample size (ESS) for binomial response in area $i$ \\
$V_{1i}$ & Estimated design-based variance on the standardized log scale for area $i$, treated as known \\
$x_{1i}$, $x_{2i}$ & Covariate vectors for the two outcomes \\
$\beta_1$, $\beta_2$ & Fixed effects for Gaussian and binomial means \\
$A$ & $N \times N$ binary adjacency matrix \\
$\Phi$ & $N \times r$ spatial basis matrix extracted from $A$ \\
$\eta$ & Shared spatial effect inducing cross-response dependence \\
$\kappa$ & Binomial-specific spatial effect capturing residual variation \\
$\tau_1, \tau_2, \tau_3$ & Coefficients scaling the shared and binomial-specific spatial effects \\
$\zeta_{1i}$, $\zeta_{2i}$ & Unstructured area random effect \\
$\omega_i$ & Latent P\'olya-Gamma variable for area $i$ \\
\hline
\end{tabular}
\end{table}

\subsection{Posterior Inference}
For the binomial block, we use the P\'olya--Gamma data augmentation \citep{polson2013bayesian}. Conditional on the latent PG variables, the logistic likelihood is Gaussian in the linear predictor, which leads to conjugate Gibbs updates.
Consider a generic binomial observation $y \sim \text{Binomial}(n,p)$ with $\text{logit}(p)=\psi$. The likelihood is proportional to
\begin{equation*}
\left\{\frac{e^\psi}{1+e^\psi}\right\}^y \left\{\frac{1}{1+e^\psi}\right\}^{n-y} = \frac{e^{y\psi}}{(1+e^\psi)^n}.
\end{equation*}
If $\omega \sim PG(b,0)$, then for any $\psi \in \mathbb{R}$ and $a \in \mathbb{R}$, the integral identity is
\begin{equation*}
\frac{e^{a\psi}}{(1+e^\psi)^b} = 2^{-b}e^{\tilde{\kappa}\psi} \mathbb{E}_\omega \left[\exp\left(-\frac{1}{2}\omega\psi^2\right)\right],
\end{equation*}
where $\tilde{\kappa} = a - \frac{b}{2}$. The corresponding conditional distribution is $\omega|\psi \sim PG(b,\psi)$.

With the {{binomial likelihood based on the effective sample size}} above, we introduce latent variables $\omega_i \mid \psi_i \sim PG(m_i^E,\psi_i)$ for $i=1,\dots,N$. Here $m_i^E$ need not be an integer. The P\'olya--Gamma representation remains valid for any positive shape parameter, and the {\tt BayesLogit} function \texttt{rpg()} allows non-integer shape inputs, so using $PG(m_i^E,\psi_i)$ is consistent with the working binomial likelihood above. This yields closed-form Gibbs updates for the model blocks. We then sample from the posterior using Gibbs sampling.

Posterior computation is carried out by Gibbs sampling using the Gaussian representation induced by the P\'olya--Gamma augmentation. Let
\[
Z_1 = (Z_{11}, \dots, Z_{1N})^\top,
\qquad
\xi_i = Z_{2i}^E - \frac{m_i^E}{2}, \quad i=1,\dots,N,
\]
and define
\[
\xi = (\xi_1,\dots,\xi_N)^\top,
\qquad
V = \mathrm{Diag}(V_{11},\dots,V_{1N}),
\qquad
\Omega = \mathrm{Diag}(\omega_1,\dots,\omega_N).
\]
Also define
\[
Q_G = \tau_1^2 \Phi^\top V^{-1}\Phi,
\qquad
Q_B = \tau_2^2 \Phi^\top \Omega \Phi.
\]
Conditional on all other parameters, the shared spatial effect has the full conditional distribution
\begin{equation*}
\eta \mid \cdot \sim N(\mu_\eta, \Sigma_\eta),
\end{equation*}
where
\begin{align*}
\Sigma_\eta &= (Q_G + Q_B + I_r)^{-1}, \\
\mu_\eta &=
\Sigma_\eta
\left[
\tau_1 \Phi^\top V^{-1}(Z_1 - X_1\beta_1 - \zeta_1)
+
\tau_2 \Phi^\top \left\{
\xi - \Omega \left(X_2\beta_2 + \tau_3 \Phi\kappa + \zeta_2\right)
\right\}
\right].
\end{align*}
The posterior covariance matrix \(\Sigma_\eta\) combines information from both the Gaussian and binomial components through the shared spatial effect. The remaining full conditional distributions are updated in the same Gibbs sampler and are provided in the Appendix.
\section{Simulation Study}\label{sec:simulation}

We conduct an empirical simulation study using 2021 ACS tract-level 5-year period estimates for South Dakota. As in earlier empirical SAE studies, we treat the ACS tract-level estimates as the ``true" values and generate noisy ``direct" estimates with sampling variability consistent with the observed ACS values; see \citet{bradley2020bayesian,parker2023computationally} for related empirical simulation designs.

{{We consider two responses. The Gaussian response is tract-level median household income. Let $M_i$ denote the ACS estimate for tract $i$, and let $\mathrm{MOE}(M_i)$ denote its published margin of error (MOE, 90\% confidence interval). The corresponding standard error is
\begin{equation*}\label{eq:se-moe-income}
  \widehat{\mathrm{SE}}(M_i)
  =
  \frac{\mathrm{MOE}(M_i)}{1.645},
\end{equation*}
and, using the delta-method approximation, the variance on the log scale is
\begin{equation*}\label{eq:log-income-var}
  \widehat{\mathrm{Var}}\{\log(M_i)\}
  \approx
  \frac{\widehat{\mathrm{SE}}(M_i)^2}{M_i^2}.
\end{equation*}

The binomial response is the 2021 ACS 5-year period tract-level direct estimate of the poverty rate. Let $p_i^{\ast}$ denote the ACS tract-level direct estimate used as the  truth in the empirical simulation study, and let $\mathrm{MOE}(p_i^{\ast})$ denote its published margin of error. The corresponding variance is
\begin{eqnarray*}\label{eq:pov-var}
  \widehat{\mathrm{Var}}(p_i^{\ast})
  =
  \left\{
    \frac{\mathrm{MOE}(p_i^{\ast})}{1.645}
  \right\}^2.
\end{eqnarray*}
For both responses, we use the same tract-level covariate vector,
\[
x_{1i} = x_{2i} = (1,\ \mathrm{White}_i)^\top,
\]
where $\mathrm{White}_i$ is the published proportion of White residents in tract $i$.}}

\subsection{Simulation Design}\label{sec:sim-design}

Let
\[
\mu_{1i}^{\ast}=\log(M_i),
\qquad
\mu_{2i}^{\ast}=\operatorname{logit}(p_i^{\ast}),
\]
and let
\[
\sigma_{1i}^2=\widehat{\mathrm{Var}}(\mu_{1i}^{\ast}),
\qquad
\sigma_{2i}^2=\widehat{\mathrm{Var}}(\mu_{2i}^{\ast})
\]
denote the ACS design-based variances on the transformed scales. {{We generate one simulated-dataset as follows.}} For each tract $i$, we draw a correlated pair of simulated data that serves as the direct estimates by
\[
\begin{pmatrix}
\widetilde{\mu}_{1i} \\
\widetilde{\mu}_{2i}
\end{pmatrix}
\Bigg|
\begin{pmatrix}
\mu_{1i}^{\ast} \\
\mu_{2i}^{\ast}
\end{pmatrix}
\sim
N_2\!\left(
\begin{pmatrix}
\mu_{1i}^{\ast} \\
\mu_{2i}^{\ast}
\end{pmatrix},
\Sigma_i
\right),
\]
where
\[
\Sigma_i
=
\begin{pmatrix}
\sigma_{1i}^2 & \rho\,\sigma_{1i}\sigma_{2i} \\
\rho\,\sigma_{1i}\sigma_{2i} & \sigma_{2i}^2
\end{pmatrix},
\qquad \rho = 0.9.
\]
We fix $\rho=0.9$ to represent a strong cross-response dependence setting and to assess whether the joint model can borrow strength across the two responses.

For the Gaussian response, we standardize the simulated direct estimate within the empirical simulated dataset
\[
Z_{1i}
=
\frac{\widetilde{\mu}_{1i}-\bar{\widetilde{\mu}}_{1}}
{s_{\widetilde{\mu}_1}},
\qquad
\widetilde{V}_{1i}
=
\frac{\sigma_{1i}^2}{s_{\widetilde{\mu}_1}^2},
\]
where $\bar{\widetilde{\mu}}_{1}$ and $s_{\widetilde{\mu}_1}$ are the sample mean and sample standard deviation of $\widetilde{\mu}_{1}$ across tracts. The Gaussian component is fitted on this standardized log-income scale. After model fitting, we transform the Gaussian posterior draws back to the log-income scale using the simulated-dataset-specific mean and standard deviation. Gaussian MSE, coverage, and interval score are then computed on the log-income scale by comparing these back-transformed posterior summaries with $\mu_{1i}^{\ast}=\log(M_i)$.

For the binomial response, we map the simulated logit direct estimate back to the probability scale using
\[
\widehat{p}_i=\operatorname{logit}^{-1}\!\left(\widetilde{\mu}_{2i}\right),
\]
and approximate its variance by
\[
\widehat{\mathrm{Var}}(\widehat{p}_i)
=
\left\{\widehat{p}_i\bigl(1-\widehat{p}_i\bigr)\right\}^2
\sigma_{2i}^2.
\]
We then compute the effective binomial sample size and effective binomial count
\[
m_i^{E}
=
\frac{\widehat{p}_i\left(1-\widehat{p}_i\right)}
{\widehat{\mathrm{Var}}(\widehat{p}_i)},
\qquad
Z_{2i}^{E} = m_i^{E}\widehat{p}_i.
\]
For each empirical simulated dataset, we fit the Univariate model and the Multi-type model using the same covariates, the same spatial basis matrix $\Phi$, and the same prior specification. Each model is run with 1{,}000 burn-in iterations and 1{,}000 retained posterior draws for inference. Performance is evaluated by tract-level MSE, empirical coverage of 95\% credible intervals, and the 95\% interval score. We repeat the entire data-generation and model-fitting procedure 100 times and report the average performance across the 100 empirical simulated datasets. {{For the Gaussian response, these summaries are computed after back-transforming the model output from the standardized log-income scale to the unstandardized log-income scale.}}

\subsection{Empirical Simulation Results}\label{sec:sim-results}
We evaluate tract-level MSE, empirical coverage of the 95\% credible intervals, and the 95\% interval score. For an interval $[l,u]$ targeting a parameter $\theta$, the interval score of \citet{gneiting2007strictly} is given by
\begin{equation*}\label{eq:IntScore}
  S_{\alpha}^{\text{int}}(l,u;\theta)
  \;=\;
  (u-l)
  + \frac{2}{\alpha}(l-\theta)\mathbf{1}\{\theta<l\}
  + \frac{2}{\alpha}(\theta-u)\mathbf{1}\{\theta>u\},
\end{equation*}
where smaller values indicate better interval performance. We report coverage and interval scores only for the model-based estimators, since the simulation focuses on posterior intervals from the fitted models.

\begin{figure}[htbp]
  \centering
  \includegraphics[width=\textwidth]{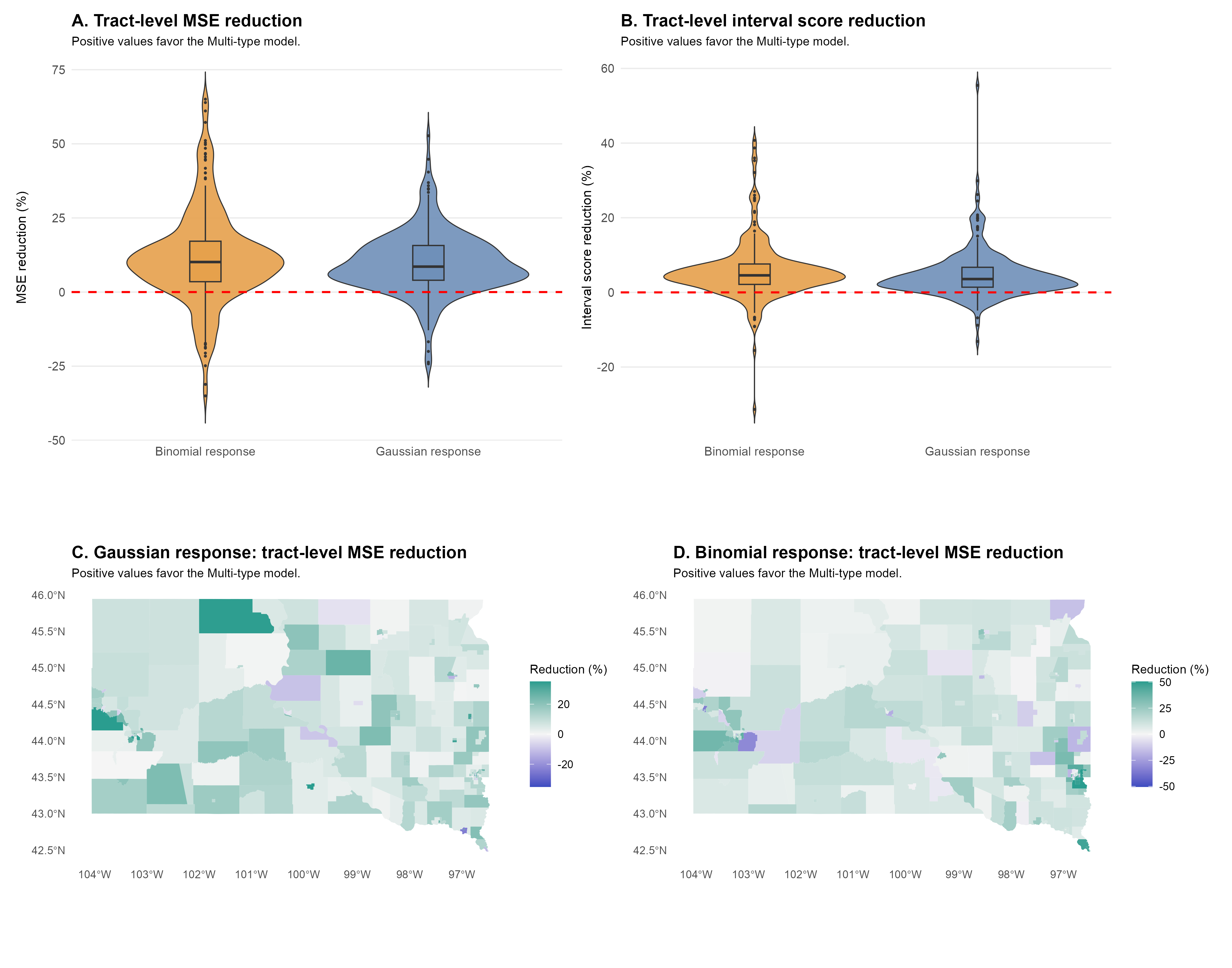}
  \caption{Empirical simulation comparison of the Univariate model and the Multi-type model. Panels A and B show tract-level percentage reductions in MSE and interval score relative to the Univariate model. Panels C and D map the tract-level MSE reduction for the Gaussian and binomial responses. Positive values favor the Multi-type model.}
  \label{fig:sim-four-panel}
\end{figure}

To compare the Univariate model and the Multi-type model directly, Figure~\ref{fig:sim-four-panel} reports tract-level percentage reductions relative to the Univariate model
\[
\mathrm{Red}_i(L)
=
100\times
\frac{L_{i,\text{Univariate}} - L_{i,\text{Multi-type}}}{L_{i,\text{Univariate}}},
\]
where $L$ is either the tract-level MSE or the tract-level interval score. Positive values favor the Multi-type model. Panels A and B show the distributions of tract-level reductions in MSE and interval score. Panels C and D map the tract-level MSE reduction for the Gaussian and binomial responses.

{{Tables~\ref{tab:gauss-sim} and \ref{tab:binom-sim} report the overall numerical summaries. For the Gaussian response, evaluated on the log-income scale, the Multi-type model further reduces both MSE and interval score relative to the Univariate model. In particular, the MSE is reduced by about 10\% compared with the Univariate model. For the binomial response, the MSE reduction from the Univariate model to the Multi-type model is about 11.77\%. Overall, for both responses, the Multi-type model achieves about a 10\% additional MSE reduction together with a smaller interval score. These results indicate that the Multi-type model improves on the Univariate model, suggesting that the cross-response borrowing strategy is appropriate. To examine tract-level performance in more detail, we turn to Figure~\ref{fig:sim-four-panel}.}}

\begin{table}[htbp]
\centering
\begin{tabular}{lcccc}
\hline
Type & MSE ($\times 10^{-3}$) & Coverage & IS & MSE Red (\%) \\
\hline
Direct estimate  & 15.5 & --    & --    & -- \\
Univariate model & 11.2 & 93.9\% & 0.449 & 27.62\% \\
Multi-type model & 9.7  & 94.8\% & 0.416 & 37.64\% \\
\hline
\end{tabular}
\caption{Empirical simulation results for the Gaussian response. MSE is reported in units of $10^{-3}$ on the log-income scale. IS denotes interval score. Reductions are relative to the direct estimate.}
\label{tab:gauss-sim}
\end{table}

\begin{table}[htbp]
\centering
\begin{tabular}{lcccc}
\hline
Type & MSE ($\times 10^{-3}$) & Coverage & IS & MSE Red (\%) \\
\hline
Direct estimate  & 5.25 & --    & --    & -- \\
Univariate model & 1.08 & 92.7\% & 0.144 & 79.50\% \\
Multi-type model & 0.96 & 93.4\% & 0.135 & 81.66\% \\
\hline
\end{tabular}
\caption{Empirical simulation results for the binomial response. MSE is reported in units of $10^{-3}$. IS denotes interval score. Reductions are relative to the direct estimate.}
\label{tab:binom-sim}
\end{table}
Figure~\ref{fig:sim-four-panel} provides a visual indication of the improvement from the Multi-type model at the tract level. Panels A and B show the distributions of tract-level percentage reductions in MSE and interval score relative to the Univariate model, and Panels C and D show the spatial patterns of tract-level MSE reduction for the Gaussian and binomial responses. For both responses, most tracts have positive reductions, indicating that the Multi-type model lowers MSE and interval score in most areas. The spatial plots show that this improvement is broadly distributed across the region. Together with the numerical results in Tables~\ref{tab:gauss-sim} and \ref{tab:binom-sim}, Figure~\ref{fig:sim-four-panel} shows that the Multi-type model provides a clear overall improvement over the Univariate model.

\section{American Community Survey Multi-type Analysis}\label{sec:data}

We analyze the West North Central region of the United States, consisting of Iowa, Kansas, Minnesota, Missouri, Nebraska, North Dakota, and South Dakota. {{After removing 48 tracts with missing ACS estimates for income or poverty, 5,859 tracts remain for analysis, and we constructed the adjacency matrix directly on this retained set of tracts using queen contiguity (i.e., where adjacency is defined by sharing a border or a corner). These omitted tracts are few and spatially scattered, so they are unlikely to affect the regional results.}} As in the simulation study, we consider two responses: log median household income and poverty rate. For both responses, we use the same tract-level covariates: the proportion of White residents, the proportion of adults with at least a bachelor's degree, and the proportion of households receiving SNAP benefits.

\subsection{Posterior Mean Surfaces}\label{sec:region-mean}
\begin{figure}[htbp]
  \centering
  \includegraphics[width=\textwidth]{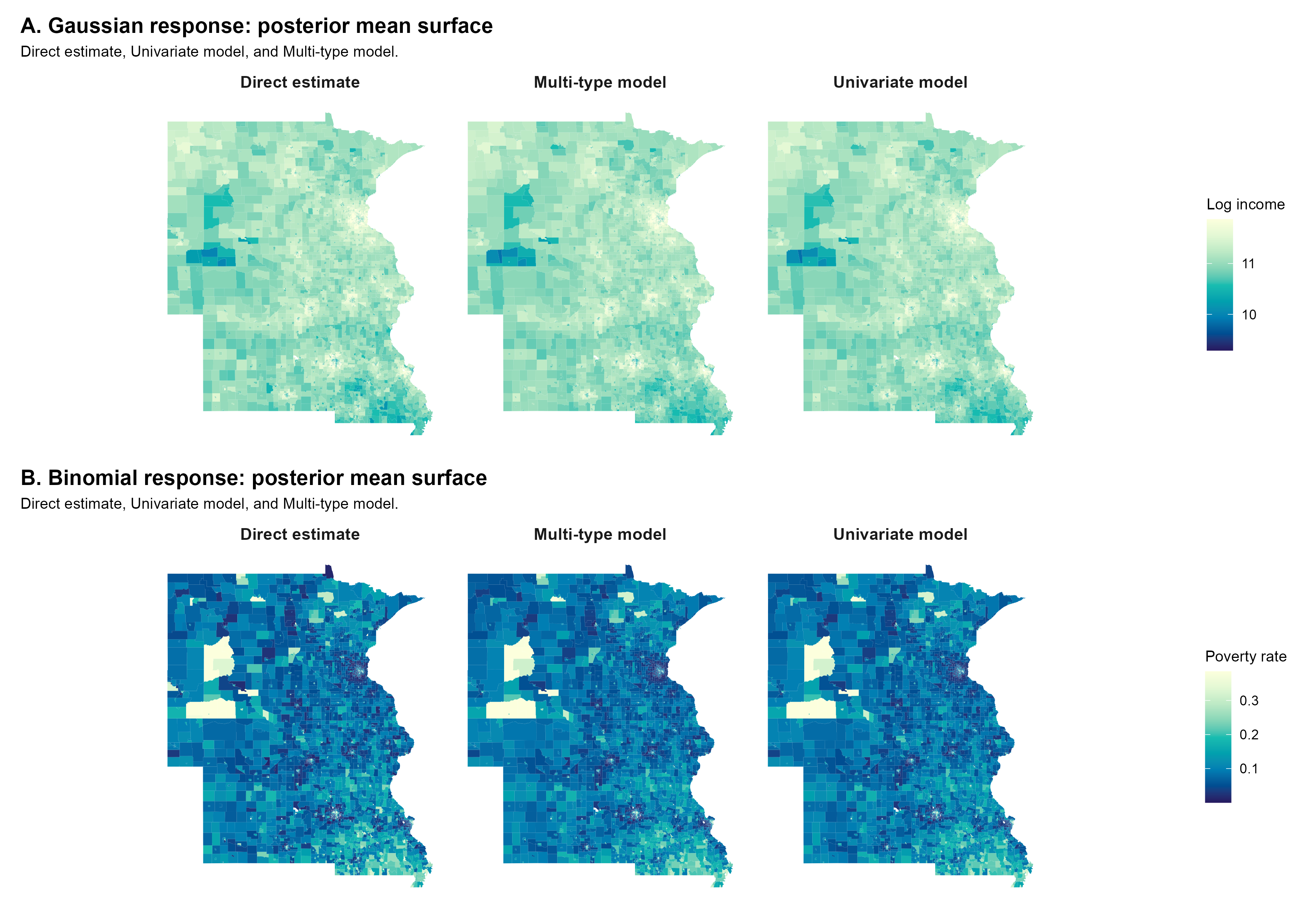}
  \caption{West North Central region: tract-level posterior mean surfaces for the Gaussian and binomial responses. Panel A shows the Gaussian response on the log-income scale, and Panel B shows the binomial response on the poverty-rate scale. In each panel, the columns show the direct estimate, the Multi-type posterior mean, and the Univariate posterior mean. Across both responses, the two model-based posterior mean surfaces remain very close to the direct estimates.}
  \label{fig:region-mean-surface}
\end{figure}

Figure~\ref{fig:region-mean-surface} compares the tract-level direct estimates, the Multi-type posterior means, and the Univariate posterior means for both responses over the West North Central region. For the Gaussian response, the three surfaces are very similar and show the same broad spatial pattern. The same is true for the binomial response. Relative to the direct estimates, both model-based surfaces are slightly smoother, but the main geographic structure is preserved.

These comparisons show that the Multi-type extension does not materially change the posterior mean surface. The main effect of the joint model therefore appears in posterior uncertainty rather than in the estimated mean pattern.

\subsection{Uncertainty Reduction}\label{sec:region-uncertainty}
\begin{figure}[htbp]
  \centering
  \includegraphics[width=1\textwidth]{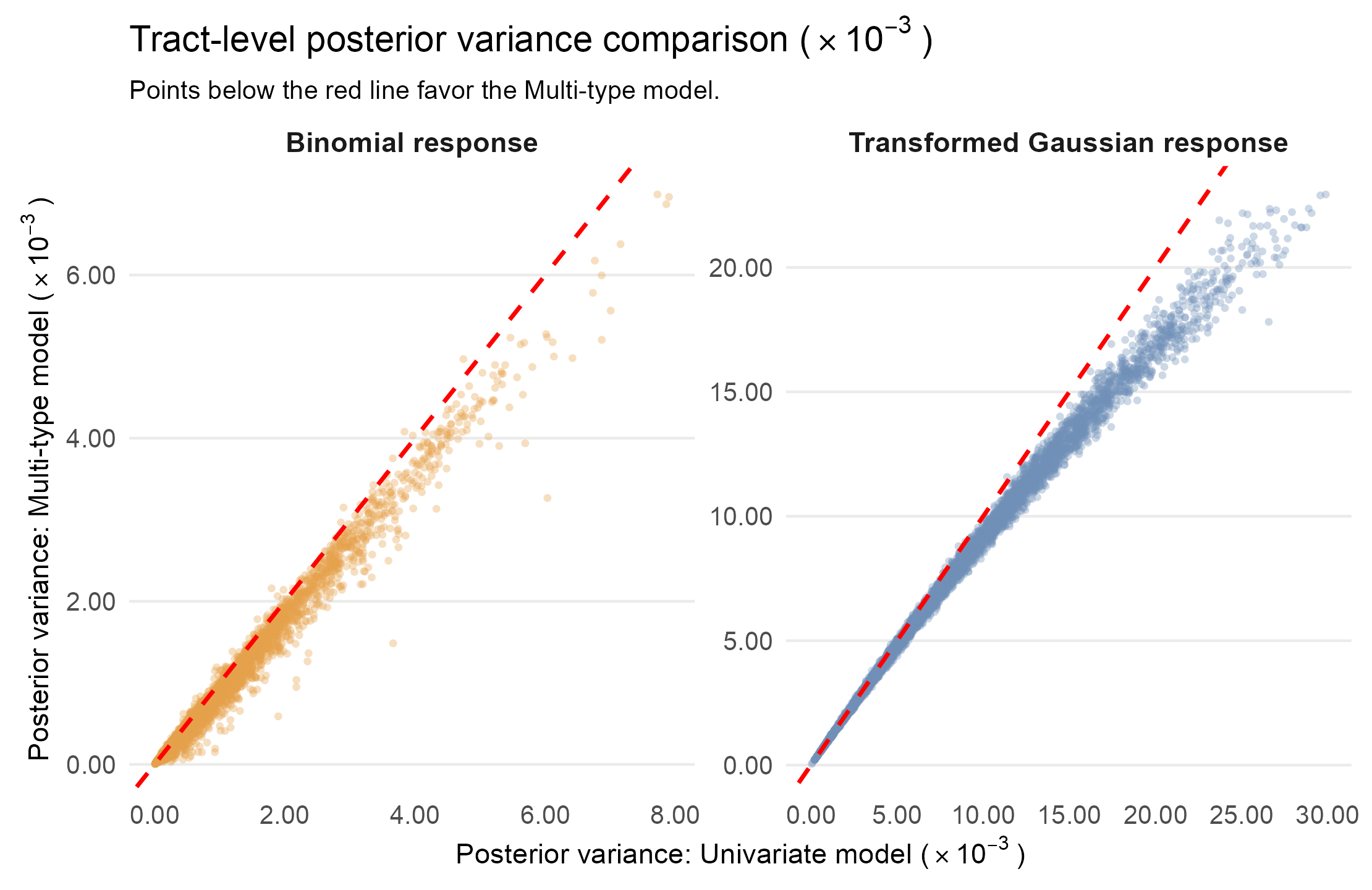}
  \caption{{{West North Central region: tract-level posterior variance comparison for the Univariate and Multi-type models. The left panel corresponds to the binomial response, and the right panel corresponds to the transformed Gaussian response. Points below the red 45-degree line favor the Multi-type model by indicating smaller posterior variance under the Multi-type model. For both responses, most tracts lie below the reference line, showing that the Multi-type model yields lower posterior uncertainty in most areas.}}}
  \label{fig:region-var-scatter}
\end{figure}

Because the posterior mean surfaces are nearly unchanged, we focus on posterior uncertainty. Figure~\ref{fig:region-var-scatter} compares tract-level posterior variances under the Univariate and Multi-type models for both responses. Points below the red 45-degree line favor the Multi-type model by indicating smaller posterior variance under the Multi-type model. For both the transformed Gaussian and binomial responses, most tracts lie below the reference line, indicating that the Multi-type model typically produces smaller posterior variances than the Univariate model.

The numerical summaries show the same pattern. For the transformed Gaussian response, the Multi-type model yields smaller posterior variance in 81.05\% of tracts. For the binomial response, the Multi-type model yields smaller posterior variance in 84.96\% of tracts. Overall, the ACS analysis shows that the direct estimates, the Univariate posterior means, and the Multi-type posterior means have very similar spatial patterns for both responses. This indicates that the Multi-type model preserves the main structure of the estimated mean surface and produces estimates that remain consistent with the direct estimates. The main difference appears in posterior uncertainty. For both responses, the Multi-type model yields smaller posterior variances for most tracts than the Univariate model. Combined with the empirical simulation results, where the Multi-type model achieves lower MSE and lower interval score, these findings indicate that the Multi-type model improves both estimation precision and uncertainty quantification.

\section{Conclusion}\label{sec:conclusion}
We propose a scalable Bayesian area-level model for jointly analyzing Gaussian and binomial survey responses. The two outcomes are linked through a shared spatial random effect, which allows the model to borrow strength across dependent responses. To retain computational efficiency and account for spatial dependence, we use P\'olya--Gamma data augmentation for the binomial component and represent the shared spatial effect through a reduced-rank basis expansion. This leads to closed-form full conditional distributions and an efficient Gibbs sampler.

In the empirical simulation study, the Multi-type model achieves lower tract-level MSE and smaller interval scores than the corresponding Univariate models for both responses. In the West North Central ACS analysis, the posterior mean surfaces from the Multi-type and Univariate models remain very close to the direct estimates, while the Multi-type model yields smaller posterior standard deviations for most tracts. Taken together, these results show that the main gain from the joint model lies in improved estimation precision and uncertainty quantification, while the estimated mean surfaces remain largely unchanged. The proposed framework provides a practical approach for area-level small area estimation with dependent Multi-type responses. The same general modeling strategy may also be extended to other non-Gaussian outcomes, such as count data and multinomial data, through appropriate augmentation schemes and likelihood formulations.

\section*{Acknowledgments}
    This article is released to inform interested parties of ongoing research and to encourage discussion. The views expressed on statistical issues are those of the authors and not those of the NSF. This research was partially supported by the U.S. National Science Foundation (NSF) under NSF grants NCSE-2215168 and NCSE-2215169.

\bibliographystyle{apalike}
\bibliography{bghm}
\section*{Appendix}\label{sec:app}
\subsection*{Notation recap and augmented likelihood}

Let $i=1,\dots,N$ index areas, and let $r$ denote the number of retained spatial basis functions. Define
\begin{eqnarray*}
\Phi &=&
\begin{bmatrix}
\phi_1^\top \\
\vdots \\
\phi_N^\top
\end{bmatrix}
\in \mathbb{R}^{N \times r},
\,\,\,\,
R = V^{-1}, \quad V = \operatorname{diag}(V_{11}, \ldots, V_{1N}),
\,\,\,\,
\Omega = \mathrm{diag}(\omega_1,\dots,\omega_N).
\end{eqnarray*}
Let
\begin{eqnarray*}
\xi_i &=& Z_{2i}^E - \frac{m_i^E}{2}, \,\,\,\,
\xi = (\xi_1,\dots,\xi_N)^\top.
\end{eqnarray*}
For the joint model, the Gaussian block is
\begin{eqnarray*}
Z_1 \mid \beta_1,\eta,\tau_1,\zeta_1 &\sim&
N(X_1\beta_1 + \tau_1 \Phi\eta + \zeta_1,\; R^{-1}),
\end{eqnarray*}
and the binomial linear predictor is
\begin{eqnarray*}
\psi &=& X_2\beta_2 + \tau_2 \Phi\eta + \tau_3 \Phi\kappa + \zeta_2.
\end{eqnarray*}
Conditional on the P\'olya--Gamma latent variables, the binomial likelihood is
\begin{eqnarray*}
p(Z_2^E \mid \beta_2,\eta,\kappa,\tau_2,\tau_3,\zeta_2,\omega)
&\propto&
\exp\left\{
\xi^\top \psi - \frac{1}{2}\psi^\top \Omega \psi
\right\}.
\end{eqnarray*}
Equivalently,
\begin{eqnarray*}
\exp\left\{
\xi^\top \psi - \frac{1}{2}\psi^\top \Omega \psi
\right\}
&\propto&
\exp\left\{
-\frac{1}{2}(\psi - \Omega^{-1}\xi)^\top \Omega (\psi - \Omega^{-1}\xi)
\right\}.
\end{eqnarray*}
The priors used in the joint model are
\begin{eqnarray*}
\beta_\ell &\sim& N(0,\sigma_\beta^{2}I_{p_\ell}), \,\,\,\, \ell=1,2,
\end{eqnarray*}
\begin{eqnarray*}
\eta &\sim& N(0,I_r), \,\,\,\, \kappa \sim N(0,I_r),
\end{eqnarray*}
\begin{eqnarray*}
\tau_j &\sim& N(0,\sigma_\tau^{2}), \,\,\,\, j=1,2,3,
\end{eqnarray*}
\begin{eqnarray*}
\zeta_\ell \mid \sigma_{\zeta_\ell}^2 &\sim& N(0,\sigma_{\zeta_\ell}^2 I_N), \,\,\,\, \ell=1,2,
\end{eqnarray*}
\begin{eqnarray*}
\sigma_{\zeta_1}^2 &\sim& IG(a_1,b_1), \,\,\,\,
\sigma_{\zeta_2}^2 \sim IG(a_2,b_2).
\end{eqnarray*}

\subsection*{Full conditional distribution of the shared spatial effect}

\textbf{Proposition 1}
Under the augmented joint model, the full conditional distribution of $\eta$ is
\begin{eqnarray*}
\eta \mid \cdot &\sim& N(Q_\eta^{-1} h_\eta,\; Q_\eta^{-1}),
\end{eqnarray*}
where
\begin{eqnarray*}
Q_\eta &=& \tau_1^2 \Phi^\top R \Phi + \tau_2^2 \Phi^\top \Omega \Phi + I_r,
\end{eqnarray*}
and
\begin{eqnarray*}
h_\eta
&=&
\tau_1 \Phi^\top R (Z_1 - X_1\beta_1 - \zeta_1)
+
\tau_2 \Phi^\top
\left[
\xi - \Omega(X_2\beta_2 + \tau_3 \Phi\kappa + \zeta_2)
\right].
\end{eqnarray*}
\textit{Proof.}
Ignoring terms not involving $\eta$, the Gaussian block contributes
\begin{eqnarray*}
-\frac{1}{2}(Z_1 - X_1\beta_1 - \tau_1 \Phi\eta - \zeta_1)^\top
R
(Z_1 - X_1\beta_1 - \tau_1 \Phi\eta - \zeta_1),
\end{eqnarray*}
the augmented binomial block contributes
\begin{eqnarray*}
\xi^\top(X_2\beta_2 + \tau_2 \Phi\eta + \tau_3 \Phi\kappa + \zeta_2)
-
\frac{1}{2}
(X_2\beta_2 + \tau_2 \Phi\eta + \tau_3 \Phi\kappa + \zeta_2)^\top
\Omega
(X_2\beta_2 + \tau_2 \Phi\eta + \tau_3 \Phi\kappa + \zeta_2),
\end{eqnarray*}
and the prior contributes
\begin{eqnarray*}
-\frac{1}{2}\eta^\top \eta.
\end{eqnarray*}
Collecting quadratic and linear terms in $\eta$ gives
\begin{eqnarray*}
\log p(\eta \mid \cdot)
&=&
-\frac{1}{2}\eta^\top Q_\eta \eta + \eta^\top h_\eta + \mathrm{const},
\end{eqnarray*}
which yields the stated multivariate normal full conditional by completing the square.
\hfill $\square$

\subsection*{Full conditional distributions for the joint model}

For convenience, define
\begin{eqnarray*}
r_{1,\beta} &=& Z_1 - \tau_1 \Phi\eta - \zeta_1,
\,\,\,\,
r_{2,\beta} = \tau_2 \Phi\eta + \tau_3 \Phi\kappa + \zeta_2,
\end{eqnarray*}
\begin{eqnarray*}
r_{1,\eta} &=& Z_1 - X_1\beta_1 - \zeta_1,
\,\,\,\,
r_{2,\eta} = \xi - \Omega(X_2\beta_2 + \tau_3 \Phi\kappa + \zeta_2).
\end{eqnarray*}
\textbf{Latent P\'olya--Gamma Variables}
\begin{eqnarray*}
\omega_i \mid \cdot &\sim& PG(m_i^E,\psi_i), \,\,\,\, i=1,\dots,N.
\end{eqnarray*}
\textbf{Fixed Effects}
\begin{eqnarray*}
\beta_1 \mid \cdot &\sim& N(Q_{\beta_1}^{-1}h_{\beta_1},\; Q_{\beta_1}^{-1}),
\,\,\,\,
Q_{\beta_1} = X_1^\top R X_1 + \sigma_\beta^{-2} I_{p_1},
\,\,\,\,
h_{\beta_1} = X_1^\top R r_{1,\beta},
\end{eqnarray*}
and
\begin{eqnarray*}
\beta_2 \mid \cdot &\sim& N(Q_{\beta_2}^{-1}h_{\beta_2},\; Q_{\beta_2}^{-1}),
\,\,\,\,
Q_{\beta_2} = X_2^\top \Omega X_2 + \sigma_\beta^{-2} I_{p_2},
\end{eqnarray*}
\begin{eqnarray*}
h_{\beta_2} &=& X_2^\top\left(\xi - \Omega r_{2,\beta}\right).
\end{eqnarray*}
\textbf{Spatial Effects}
\begin{eqnarray*}
\eta \mid \cdot &\sim& N(Q_\eta^{-1}h_\eta,\; Q_\eta^{-1}),
\end{eqnarray*}
with $Q_\eta$ and $h_\eta$ given above, and
\begin{eqnarray*}
\kappa \mid \cdot &\sim& N(Q_\kappa^{-1}h_\kappa,\; Q_\kappa^{-1}),
\end{eqnarray*}
where
\begin{eqnarray*}
Q_\kappa &=& \tau_3^2 \Phi^\top \Omega \Phi + I_r,
\end{eqnarray*}
\begin{eqnarray*}
h_\kappa
&=&
\tau_3 \Phi^\top
\left[
\xi - \Omega(X_2\beta_2 + \tau_2 \Phi\eta + \zeta_2)
\right].
\end{eqnarray*}
\textbf{Area-level Random Effect}
\begin{eqnarray*}
\zeta_1 \mid \cdot &\sim& N(Q_{\zeta_1}^{-1}h_{\zeta_1},\; Q_{\zeta_1}^{-1}),
\,\,\,\,
Q_{\zeta_1} = R + \sigma_{\zeta_1}^{-2} I_N,
\end{eqnarray*}
\begin{eqnarray*}
h_{\zeta_1} &=& R(Z_1 - X_1\beta_1 - \tau_1 \Phi\eta),
\end{eqnarray*}
and
\begin{eqnarray*}
\zeta_2 \mid \cdot &\sim& N(Q_{\zeta_2}^{-1}h_{\zeta_2},\; Q_{\zeta_2}^{-1}),
\,\,\,\,
Q_{\zeta_2} = \Omega + \sigma_{\zeta_2}^{-2} I_N,
\end{eqnarray*}
\begin{eqnarray*}
h_{\zeta_2}
&=&
\xi - \Omega(X_2\beta_2 + \tau_2 \Phi\eta + \tau_3 \Phi\kappa).
\end{eqnarray*}
\textbf{Scaling Parameters}
\begin{eqnarray*}
\tau_1 \mid \cdot &\sim& N(Q_{\tau_1}^{-1}h_{\tau_1},\; Q_{\tau_1}^{-1}),
\,\,\,\,
Q_{\tau_1} = \eta^\top \Phi^\top R \Phi \eta + \sigma_\tau^{-2},
\end{eqnarray*}
\begin{eqnarray*}
h_{\tau_1} &=& \eta^\top \Phi^\top R (Z_1 - X_1\beta_1 - \zeta_1),
\end{eqnarray*}
\begin{eqnarray*}
\tau_2 \mid \cdot &\sim& N(Q_{\tau_2}^{-1}h_{\tau_2},\; Q_{\tau_2}^{-1}),
\,\,\,\,
Q_{\tau_2} = \eta^\top \Phi^\top \Omega \Phi \eta + \sigma_\tau^{-2},
\end{eqnarray*}
\begin{eqnarray*}
h_{\tau_2}
&=&
\eta^\top \Phi^\top
\left[
\xi - \Omega(X_2\beta_2 + \tau_3 \Phi\kappa + \zeta_2)
\right],
\end{eqnarray*}
and
\begin{eqnarray*}
\tau_3 \mid \cdot &\sim& N(Q_{\tau_3}^{-1}h_{\tau_3},\; Q_{\tau_3}^{-1}),
\,\,\,\,
Q_{\tau_3} = \kappa^\top \Phi^\top \Omega \Phi \kappa + \sigma_\tau^{-2},
\end{eqnarray*}
\begin{eqnarray*}
h_{\tau_3}
&=&
\kappa^\top \Phi^\top
\left[
\xi - \Omega(X_2\beta_2 + \tau_2 \Phi\eta + \zeta_2)
\right].
\end{eqnarray*}
\textbf{Variance Parameters}
\begin{eqnarray*}
\sigma_{\zeta_1}^2 \mid \cdot &\sim&
IG\left(a_1 + \frac{N}{2},\; b_1 + \frac{1}{2}\zeta_1^\top \zeta_1\right),
\end{eqnarray*}
\begin{eqnarray*}
\sigma_{\zeta_2}^2 \mid \cdot &\sim&
IG\left(a_2 + \frac{N}{2},\; b_2 + \frac{1}{2}\zeta_2^\top \zeta_2\right).
\end{eqnarray*}

Thus, all full conditional distributions in the joint model are available in closed form.

\subsection*{Univariate Gaussian Baseline}

For the univariate Gaussian baseline, the Gaussian response is modeled separately with its own spatial coefficient vector $u_1$
\begin{eqnarray*}
Z_1 \mid \beta_1,u_1,\zeta_1 &\sim& N(X_1\beta_1 + \Phi u_1 + \zeta_1,\; R^{-1}),
\end{eqnarray*}
with priors
\begin{eqnarray*}
\beta_1 &\sim& N(0,\sigma_\beta^{2}I_{p_1}), \,\,\,\,
u_1 \mid \sigma_{u_1}^2 \sim N(0,\sigma_{u_1}^2 I_r),
\end{eqnarray*}
\begin{eqnarray*}
\zeta_1 \mid \sigma_{\zeta_1}^2 &\sim& N(0,\sigma_{\zeta_1}^2 I_N), \,\,\,\,
\sigma_{u_1}^2 \sim IG(a_u,b_u), \,\,\,\,
\sigma_{\zeta_1}^2 \sim IG(a_{\zeta_1},b_{\zeta_1}).
\end{eqnarray*}
The full conditionals are
\begin{eqnarray*}
\beta_1 \mid \cdot &\sim& N(Q_{\beta_1}^{-1}h_{\beta_1},\; Q_{\beta_1}^{-1}),
\,\,\,\,
Q_{\beta_1}=X_1^\top R X_1 + \sigma_\beta^{-2} I_{p_1},
\end{eqnarray*}
\begin{eqnarray*}
h_{\beta_1}=X_1^\top R(Z_1 - \Phi u_1 - \zeta_1),
\end{eqnarray*}
\begin{eqnarray*}
u_1 \mid \cdot &\sim& N(Q_{u_1}^{-1}h_{u_1},\; Q_{u_1}^{-1}),
\,\,\,\,
Q_{u_1}=\Phi^\top R \Phi + \sigma_{u_1}^{-2}I_r,
\end{eqnarray*}
\begin{eqnarray*}
h_{u_1}=\Phi^\top R(Z_1 - X_1\beta_1 - \zeta_1),
\end{eqnarray*}
\begin{eqnarray*}
\zeta_1 \mid \cdot &\sim& N(Q_{\zeta_1}^{-1}h_{\zeta_1},\; Q_{\zeta_1}^{-1}),
\,\,\,\,
Q_{\zeta_1}=R+\sigma_{\zeta_1}^{-2}I_N,
\end{eqnarray*}
\begin{eqnarray*}
h_{\zeta_1}=R(Z_1 - X_1\beta_1 - \Phi u_1),
\end{eqnarray*}
\begin{eqnarray*}
\sigma_{u_1}^2 \mid \cdot &\sim&
IG\left(a_u + \frac{r}{2},\; b_u + \frac{1}{2}u_1^\top u_1\right),
\end{eqnarray*}
\begin{eqnarray*}
\sigma_{\zeta_1}^2 \mid \cdot &\sim&
IG\left(a_{\zeta_1} + \frac{N}{2},\; b_{\zeta_1} + \frac{1}{2}\zeta_1^\top \zeta_1\right).
\end{eqnarray*}

\subsection*{Univariate Binomial Baseline}
For the univariate binomial baseline, the binomial response is modeled separately with its own spatial coefficient vector $u_2$
\begin{eqnarray*}
\psi &=& X_2\beta_2 + \Phi u_2 + \zeta_2,
\end{eqnarray*}
and, conditional on P\'olya--Gamma variables,
\begin{eqnarray*}
p(Z_2^E \mid \beta_2,u_2,\zeta_2,\omega)
&\propto&
\exp\left\{
\xi^\top \psi - \frac{1}{2}\psi^\top \Omega \psi
\right\}.
\end{eqnarray*}
The priors are
\begin{eqnarray*}
\beta_2 &\sim& N(0,\sigma_\beta^{2}I_{p_2}), \,\,\,\,
u_2 \mid \sigma_{u_2}^2 \sim N(0,\sigma_{u_2}^2 I_r),
\end{eqnarray*}
\begin{eqnarray*}
\zeta_2 \mid \sigma_{\zeta_2}^2 &\sim& N(0,\sigma_{\zeta_2}^2 I_N), \,\,\,\,
\sigma_{u_2}^2 \sim IG(a_u,b_u), \,\,\,\,
\sigma_{\zeta_2}^2 \sim IG(a_{\zeta_2},b_{\zeta_2}).
\end{eqnarray*}
The full conditionals are
\begin{eqnarray*}
\omega_i \mid \cdot &\sim& PG(m_i^E,\psi_i), \,\,\,\, i=1,\dots,N,
\end{eqnarray*}
\begin{eqnarray*}
\beta_2 \mid \cdot &\sim& N(Q_{\beta_2}^{-1}h_{\beta_2},\; Q_{\beta_2}^{-1}),
\,\,\,\,
Q_{\beta_2}=X_2^\top \Omega X_2 + \sigma_\beta^{-2} I_{p_2},
\end{eqnarray*}
\begin{eqnarray*}
h_{\beta_2}=X_2^\top\left(\xi - \Omega(\Phi u_2 + \zeta_2)\right),
\end{eqnarray*}
\begin{eqnarray*}
u_2 \mid \cdot &\sim& N(Q_{u_2}^{-1}h_{u_2},\; Q_{u_2}^{-1}),
\,\,\,\,
Q_{u_2}=\Phi^\top \Omega \Phi + \sigma_{u_2}^{-2}I_r,
\end{eqnarray*}
\begin{eqnarray*}
h_{u_2}=\Phi^\top\left(\xi - \Omega(X_2\beta_2 + \zeta_2)\right),
\end{eqnarray*}
\begin{eqnarray*}
\zeta_2 \mid \cdot &\sim& N(Q_{\zeta_2}^{-1}h_{\zeta_2},\; Q_{\zeta_2}^{-1}),
\,\,\,\,
Q_{\zeta_2}=\Omega+\sigma_{\zeta_2}^{-2}I_N,
\end{eqnarray*}
\begin{eqnarray*}
h_{\zeta_2}=\xi - \Omega(X_2\beta_2 + \Phi u_2),
\end{eqnarray*}
\begin{eqnarray*}
\sigma_{u_2}^2 \mid \cdot &\sim&
IG\left(a_u + \frac{r}{2},\; b_u + \frac{1}{2}u_2^\top u_2\right),
\end{eqnarray*}
\begin{eqnarray*}
\sigma_{\zeta_2}^2 \mid \cdot &\sim&
IG\left(a_{\zeta_2} + \frac{N}{2},\; b_{\zeta_2} + \frac{1}{2}\zeta_2^\top \zeta_2\right).
\end{eqnarray*}
\end{document}